# Selective epitaxial growth of GaAs on Ge by MOCVD

Guy Brammertz[*], Yves Mols, Stefan Degroote, Maarten Leys, Jan Van Steenbergen, Gustaaf Borghs, and Matty Caymax

Interuniversity Microelectronics Center (IMEC vzw), Kapeldreef 75, B-3001 Leuven, Belgium.

**Abstract**
We have selectively grown thin epitaxial GaAs films on Ge substrates with the aid of a 200 nm thin $SiO_2$ mask layer. The selectively grown structures have lateral sizes ranging from 1 µm width up to large areas of 1 by 1 $mm^2$. The growth with the standard growth procedure for GaAs growth on Ge substrates reveals a limited amount of GaAs nucleation on the mask area and strong loading effects caused by diffusion of group III precursors over the mask area and in the gas phase. Reduction of the growth pressure inhibits GaAs nucleation on the mask area and reduces the loading effects strongly, but favors the creation of anti phase domains in the GaAs. An optimized growth procedure was developed, consisting of a 13 nm thin nucleation layer grown at high pressure, followed by low pressure growth of GaAs. This optimized growth procedure inhibits the nucleation of GaAs on the mask area and is a good compromise between reduction of loading effects and inhibition of anti phase domain growth in the GaAs. X-ray diffraction and photoluminescence measurements demonstrate the good microscopic characteristics of the selectively grown layers.

---

[*] Corresponding author. Tel.: +32 16 28 8120; fax: +32 16 28 1214.

*Email address*: Guy.Brammertz@imec.be

# 1. Introduction

As Si device scaling for future generations of complementary metal-oxide semiconductor (CMOS) circuits becomes increasingly difficult [1], an alternative option becomes more appealing: GaAs and Ge are intrinsically faster semiconductors than Si, because of their higher bulk mobilities [2]. Especially the integration of Ge PMOS with GaAs NMOS is a very attractive replacement for fast Si CMOS structures [3]. In addition, with the introduction of high-k materials for gate insulation, a clear advantage of Si CMOS has vanished: the easy surface passivation by $SiO_2$. For that reason, a growing research interest in GaAs [4,5] and Ge [6,7] MOS transistor structures has been observed recently. For the co-integration of GaAs and Ge MOS structures in a planar scalable technology, thin films of GaAs and Ge will have to reside side-by-side on the very same substrate. The easiest way to achieve this goal is selective epitaxial growth of thin GaAs layers on Ge substrates. Even though selective growth of GaAs on GaAs substrates [8-17] and on Si substrates [18-21] has already been studied extensively, very little is known about selective growth of GaAs on Ge.

In general, growth of GaAs on Ge presents a somewhat increased complexity compared to the GaAs substrate case, because of the polar nature of GaAs, compared to the non-polar nature of the Ge substrates. This necessitates the introduction of a Ge substrate treatment and the growth of a high pressure nucleation layer in order to suppress the formation of anti phase domains (APDs) in the GaAs [22-24]. The boundaries between the different APDs consist of electrically very active Ga-Ga and As-As bonds, which act as strong scattering centers and inhibit the free charge carrier flow with high mobility.

On the other hand, GaAs growth on Ge substrates presents a serious advantage over the growth on Si substrates, because of the almost identical lattice constant of GaAs and Ge. Si presents a 4% lattice mismatch compared to GaAs.

In the following, we present the results of a study of GaAs selective growth on Ge substrates, with a thin $SiO_2$ mask layer. In the first paragraph we describe the metalorganic chemical vapor deposition (MOCVD) tool, in which the growths were performed, followed by a short description of the Ge substrate and $SiO_2$ mask layer. We then describe the selectively grown layers using the standard growth procedure for GaAs growth on Ge substrates. Subsequently, low pressure selective growth on Ge is inspected and followed by the presentation of an optimized growth procedure for selective growth of GaAs on Ge substrates. Finally, the material quality of the selectively grown GaAs is inspected with x-ray diffraction (XRD) and photoluminescence (PL) spectroscopy.

## 2. Experimental set-up

The GaAs deposition tool is a Thomas Swan MOCVD reactor. The deposition chamber consists of a vertical flow reactor with a 3-zone carbon heater, able to reach a maximum temperature of about 800°C, and a silicon carbide (SiC) coated carbon susceptor, which holds the 100 mm substrates on which we perform the growths. The reactants impinge on the susceptor from the top via a close-coupled showerhead, which distributes the gas flow homogeneously over the surface of the wafer. From there the gases are vertically removed from the reactor through the side of the susceptor via a quartz liner system. The total pressure in the reactor can be varied from 30 torr to atmospheric pressure.

The precursor sources for the MOCVD growth are Trimethylgallium (TMGa) and Tertiarybutylarsine (TBAs). The $H_2$ carrier gas flow for all experiments is 16 standard liters per minute (slm).

The GaAs analyzed in this report is grown on bulk 100 mm Ge (001) wafers acquired from UMICORE. For APD-free growth, the substrate needs to have a misalignment with respect to the exact [001] direction. In our case we chose for a misalignment of 6° towards the (111) direction, which allows for a sufficient density of single atomic steps on the Ge surface. On these wafers a 200 nm thick amorphous $SiO_2$ film is deposited and subsequently patterned and etched selectively in an HF solution, in order to reveal the Ge in the holes of the dielectric. The maskset chosen is a general mask used for transistor processing and contains structures as small as a micron width up to larger structures of a mm square, allowing for a very large spread in feature sizes and filling factor of the $SiO_2$ over the area of the mask. The global filling factor, which is the ratio of open area to the total area of the mask, is equal to 40%, which means that the larger part of the area is covered with $SiO_2$. Loading effects, caused by diffusion of growth species on the mask surface and desorption of growth species from the mask into the gas phase, should therefore be important, allowing a good study of the effects of the selective growth. Figure 1 shows an overview of the mask, highlighting the different feature sizes. From the layout of the mask it becomes clear that quantitative characterization of the loading effects is very difficult with this layout. We will therefore focus on a qualitative analysis of the loading effect. A dedicated maskset is needed for a more systematic study of the loading effects. For all growths, a substrate temperature of 660°C was adapted, as this temperature showed the best compromise between desorption of the group III precursor from the $SiO_2$, while not decomposing the group V precursor on the $SiO_2$, allowing for the best selective growth of GaAs with a $SiO_2$ mask [8].

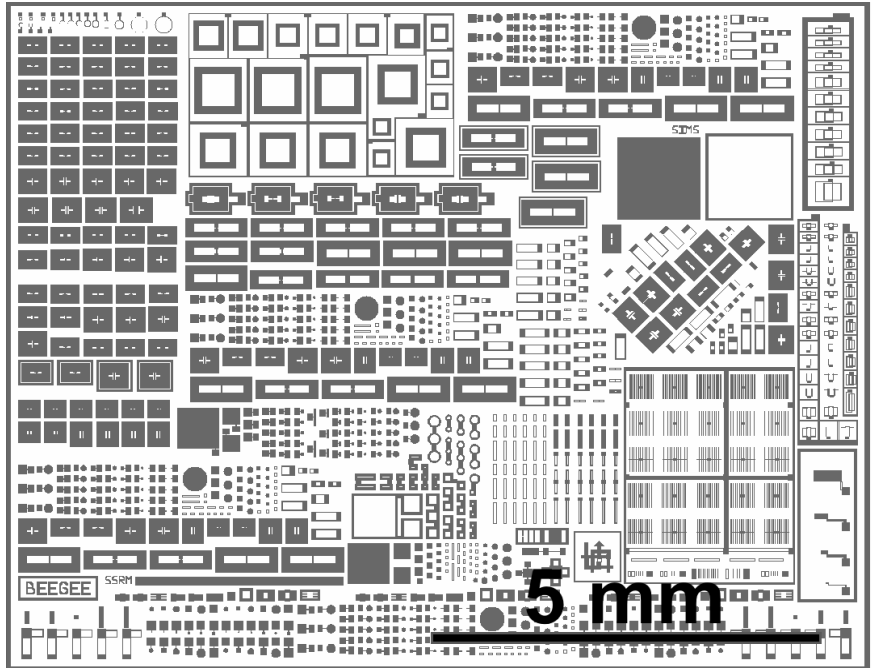

**Figure 1:** Schematic representation of the maskset used for selective epitaxial growth. The white area represents the SiO$_2$ mask, whereas the black area is open towards the Ge substrate.

### 3. Standard growth procedure

The regular growth procedure for GaAs on Ge consists of a bake of the Ge substrate at high temperature, allowing single atomic surface steps of the 6° miscut (001) surface to merge into energetically more stable double surface steps. This bake is followed by a flush of the substrate with a large TBAs flow, allowing for a complete, self-limiting coverage of the Ge surface with As. Relatively high pressure and a large TBAs flow are necessary in order to maximize the As partial pressure on top of the substrate, which assures the full monolayer coverage of the Ge, without desorption of the very volatile As from the surface. After the growth of 35 nm of GaAs with an As partial pressure of 25 Torr, a total pressure of 450 Torr and a V/III ratio of about 100, all three parameters are ramped down to respectively one Torr, 76 Torr and 16. At the end of the ramp, the grown nucleation layer is about 100 nm thick. Now growth can be continued with the standard growth pressure of 76 Torr, a growth rate of 1 nm/sec and a V/III ratio equal to 16, allowing for the growth of very high quality GaAs on Ge.
This standard growth procedure was adapted for GaAs growth on the SiO$_2$-patterned Ge wafers. The total growth time was chosen in order to achieve a total growth thickness of 300 nm on an unpatterned Ge wafer. Figure 2 shows a Nomarski microscope image (a), a cross-section scanning electron microscopy (SEM) picture (b) and a height profilometer measurement (c) of the selectively grown layer. The Nomarski micrograph shows that selective growth is possible even though a limited amount of GaAs nucleation takes place on the SiO$_2$ mask area, as can be witnessed from the small black spots present on the dielectric. The cross-section SEM picture shows a detail of a selectively grown 12 µm wide GaAs stripe. The stripe is at an angle of 45° with respect to the cleavage plane, explaining the larger width of the structure and the presence of GaAs towards the left hand side of the picture, which is the GaAs in the background of the sample. A strong microscopic loading effect can be seen in the picture. Even though the thickness was planned to be only 300 nm on a planar substrate, the thickness of the selectively grown structure in the SEM picture is 1.5 µm. Also a thickness enhancement of the film can be seen towards the interface with the SiO$_2$ layer. These loading effects arise from the diffusion of the group III precursor on the surface of the SiO$_2$ and desorption of the group III precursor from the surface of the SiO$_2$, resulting in a higher concentration of group III precursor impinging on the opening in the dielectric layer [17]. The actual concentration of the group III precursor depends strongly on the local filling factor of the mask, which is the ratio of the open mask area to the total area [17]. The profilometer measurement also shows these loading effects very clearly, visualizing the typical U-shaped surfaces of the selectively grown structures. A consequence of these loading effects is that the thickness of the grown GaAs islands varies strongly, with the largest islands surrounded by the least mask area being the thinnest and the smallest islands surrounded by the largest mask area being the thickest. The GaAs thickness varies from 600 nm to about 2 µm, depending on the local filling factor. Similarly, a larger thickness enhancement towards the edge of a GaAs island is observed for a larger adjacent mask area, whereas a smaller thickness enhancement corresponds to a smaller adjacent mask area. The spikes on the left hand side of figure 2c are small GaAs crystals, which nucleate on larger SiO$_2$ areas.

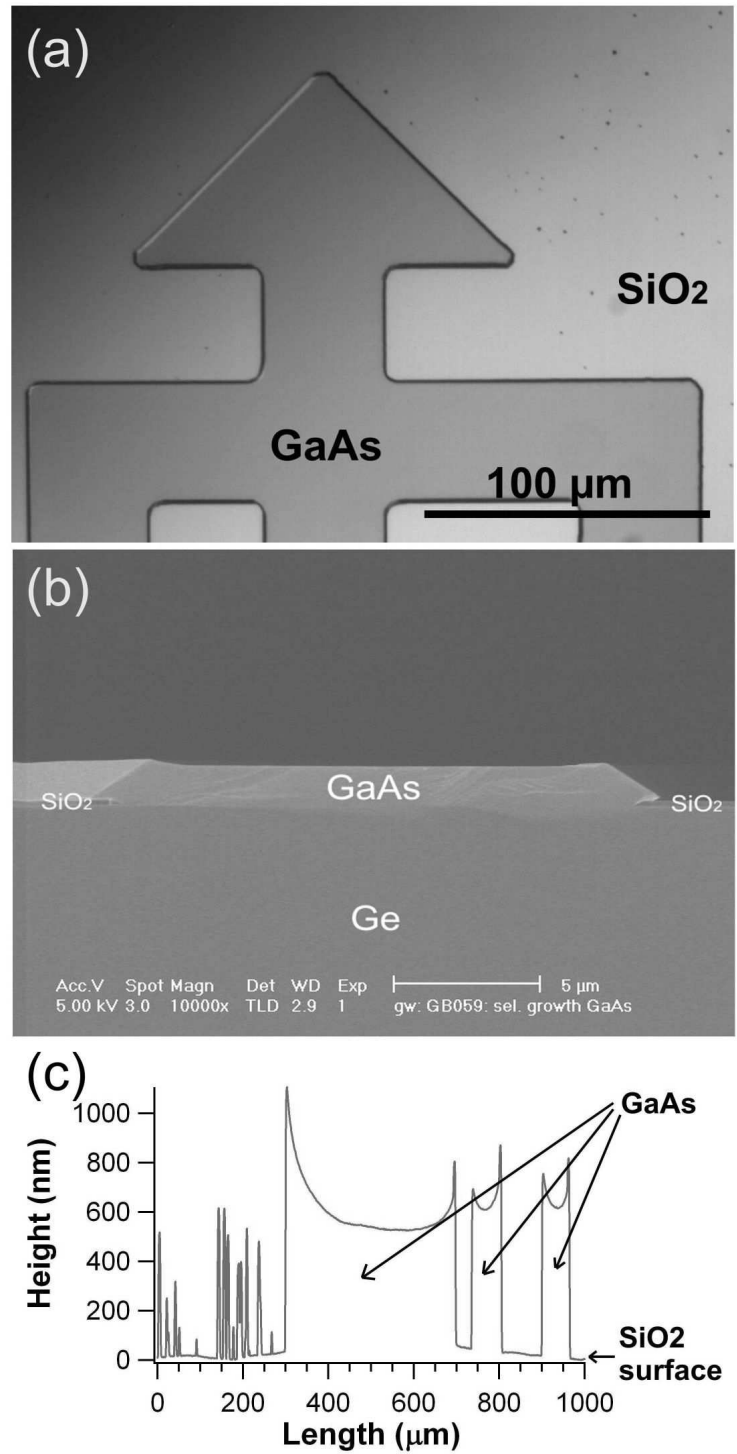

**Figure 2:** Nomarski microscope image (a), cross section SEM picture (b) and profilometer measurement (c) of a selectively grown GaAs layer on Ge, grown with the standard growth procedure.

Previous studies on MOCVD selective growth of GaAs on GaAs and Si substrates have shown two main solutions to the loading effect problem. The first solution is low pressure MOCVD (LP-MOCVD), increasing the mean free path of the species in the gas phase and accordingly increasing the speed with which the precursors are removed from the mask area [9,12,15]. This reduces diffusion of the group III precursor on the surface of the mask area, as well as diffusion of desorbed species through the gas phase and redeposition in the adjacent openings of the mask. A second solution to the selective growth problem is growth in a Chlorine environment, either by adding HCl to the carrier gas [11,13] or by using a Cl-containing precursor [10,15]. The presence of Cl in the growth environment assures the creation of very volatile and stable chlorides, increasing the desorption of group III species from the substrate and inhibiting the redeposition of these desorbed species in the openings of the mask.

In the following we will further inspect how the reduction of growth pressure influences the characteristics of the selectively grown GaAs on Ge.

## 4. Low pressure growth

In order to reduce the strong loading effects and nucleation of GaAs on the mask area, as witnessed for the standard growth procedure, we have grown GaAs on Ge at low pressure. While leaving all other growth parameters the same as for the standard growth procedure, the full growth of the GaAs film on Ge was performed at 30 Torr. The growth time was chosen in order to have a total GaAs film thickness of 300 nm on an unpatterned Ge substrate.

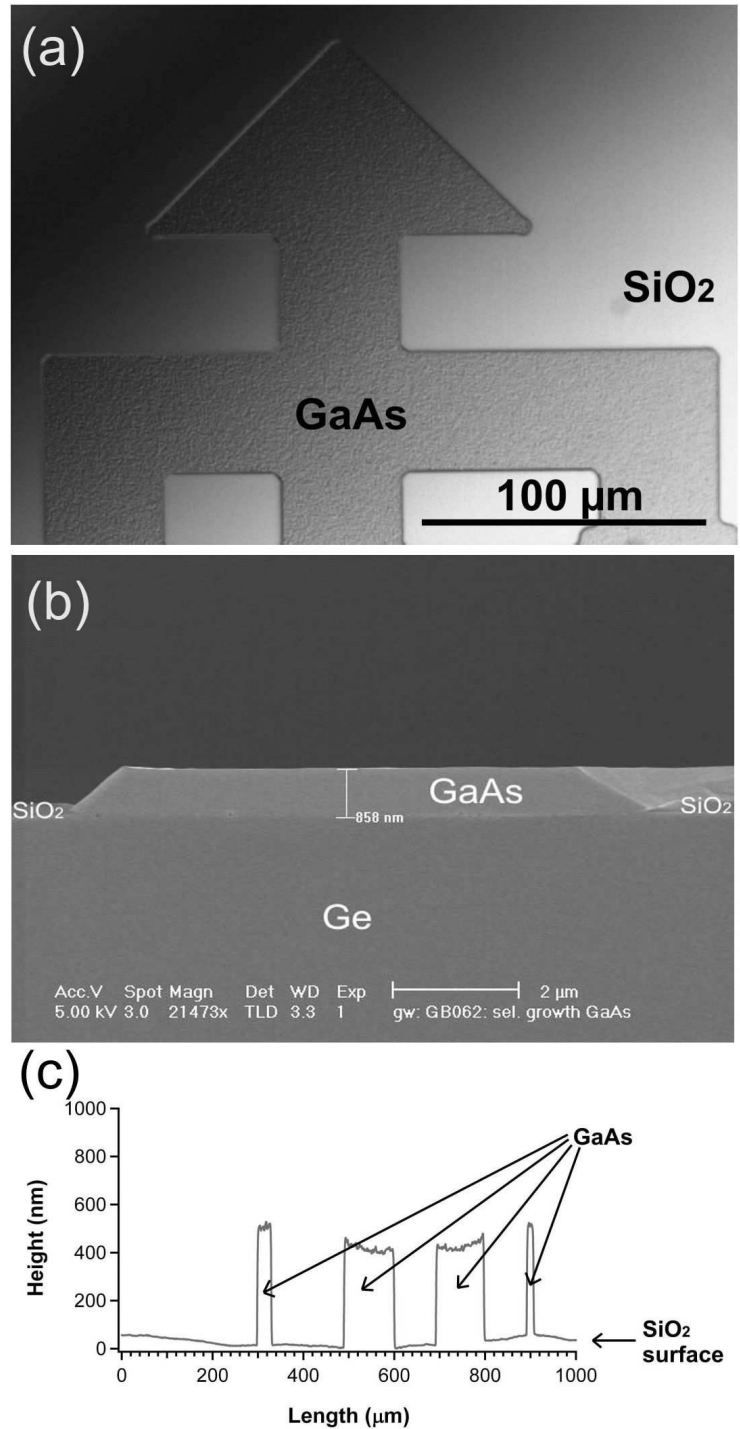

**Figure 3:** Nomarski microscope image (a), cross section SEM picture (b) and profilometer measurement (c) of a selectively grown GaAs layer on Ge, grown with the low pressure growth procedure.

Figure 3 shows a Nomarski microscope image (a), a cross-section SEM picture (b) and a height profilometer measurement (c) of the layer grown selectively at low pressure. The Nomarski microscope picture shows that selectivity is very good, with no GaAs nucleation on the $SiO_2$, even for the largest 1x1 mm$^2$ structures. Nevertheless, the morphology of the GaAs is very bad, with a large surface roughness that can easily be identified on the Nomarski picture. This large surface roughness is probably caused by a high density of APDs, which arise from bad nucleation on the Ge, because of the low pressure. The low pressure during the initial nucleation step causes the initial monolayer of As coverage of the substrate to desorb, causing bad quality GaAs to grow with a high density of APDs. These APDs cause a local reduction of the growth rate at the boundary between different domains, causing the formation of a V-shaped groove at the surface of the layer. This is illustrated in figure 4, which shows a transmission electron microscopy (TEM) picture of a 300 nm planar GaAs film grown on Ge with APDs. On the other hand, the cross section SEM picture of a selectively grown 3 µm wide stripe shows that loading effects are strongly reduced, with the layer being almost flat at the interfaces with the $SiO_2$. Again the structure is inclined 45° with respect to the cleavage plane, explaining the larger width on the picture, as well as the presence of GaAs on the right hand side of the image, which is due to GaAs in the background of the sample. A closer inspection of the interface also reveals a certain number of voids at the Ge/GaAs heterointerface, some of them having a size of almost 50 nm in diameter. These voids are probably caused by incomplete coverage of the Ge with As and the consequently bad nucleation properties of the low pressure GaAs on Ge. The profilometer measurement confirms the small loading effects of the layer, with all GaAs islands being reasonably flat and of equal height. On the other hand, the bad surface morphology can be seen on the profilometer measurement as well.

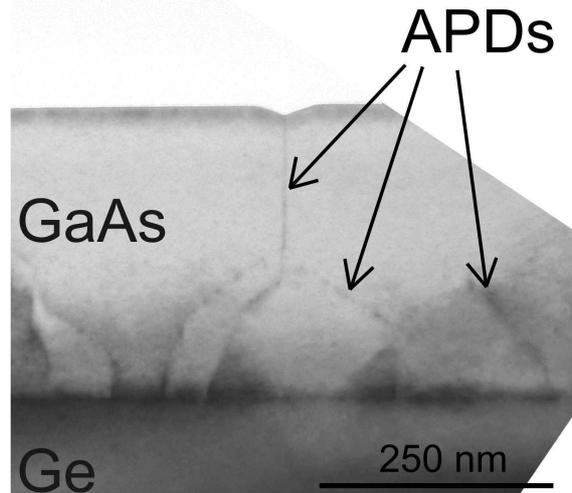

**Figure 4:** TEM picture of a thin GaAs layer containing anti phase domains grown on a Ge substrate. The boundaries between the different APDs are highlighted.

## 5. Optimized growth procedure

Even though the loading effects are strongly reduced for GaAs layers grown at low pressure, the complete growth at low pressure on Ge substrates is not advisable, because of the creation of APDs in the layer. The boundaries between the different APDs are highly charged structures in the GaAs crystal and act as strong scattering centers. Mobility and carrier lifetimes are strongly reduced because of these boundaries, and they must in all cases be avoided for high quality electronic or optoelectronic applications. For this reason we have developed a growth procedure, which consists of a minimized growth at high pressure for the very first nucleation step, followed by a fast ramp-down to lower pressure and subsequent low pressure growth. The procedure also includes the reduction of the growth rate of the very first nucleation layer to about 0.1 nm per second. About 5 nm of GaAs is grown with an As partial pressure of 25 Torr, a total reactor pressure of 450 Torr and a V/III ratio of about 100. Subsequently, all three parameters are ramped down to values of 1 Torr, 30 Torr and 16 respectively, during which another 8 nm of GaAs is deposited. Then, the residual layer is deposited with these conditions and with a growth rate of 0.5 nm/sec. The nucleation layer is therefore only 13 nm thick, just enough for the creation of a continuous film of GaAs on the Ge substrate and thin enough to avoid nucleation of GaAs on the $SiO_2$. Figure 5 shows respectively a Nomarski microscope picture (a), a cross-section SEM image (b) and a profilometer measurement (c) of a selectively grown layer with this optimized growth procedure. The total growth time was chosen in order to achieve a 150 nm thick film on an unpatterned substrate. The Nomarski microscope picture shows that the growth is fully selective. Even on the largest 1x1 $mm^2$ $SiO_2$ areas no GaAs is nucleated. The surface morphology of the layer is much better than for the low pressure growth. The cross section SEM picture of a 12 µm wide stripe shows that the loading effect is reduced as compared to the growth procedure with the thick nucleation layer. Nevertheless, from the profilometer measurement, it becomes apparent that the typical U-shaped nature of the selectively grown structures is still present, although to a lesser extent than compared to the standard nucleation layer case.

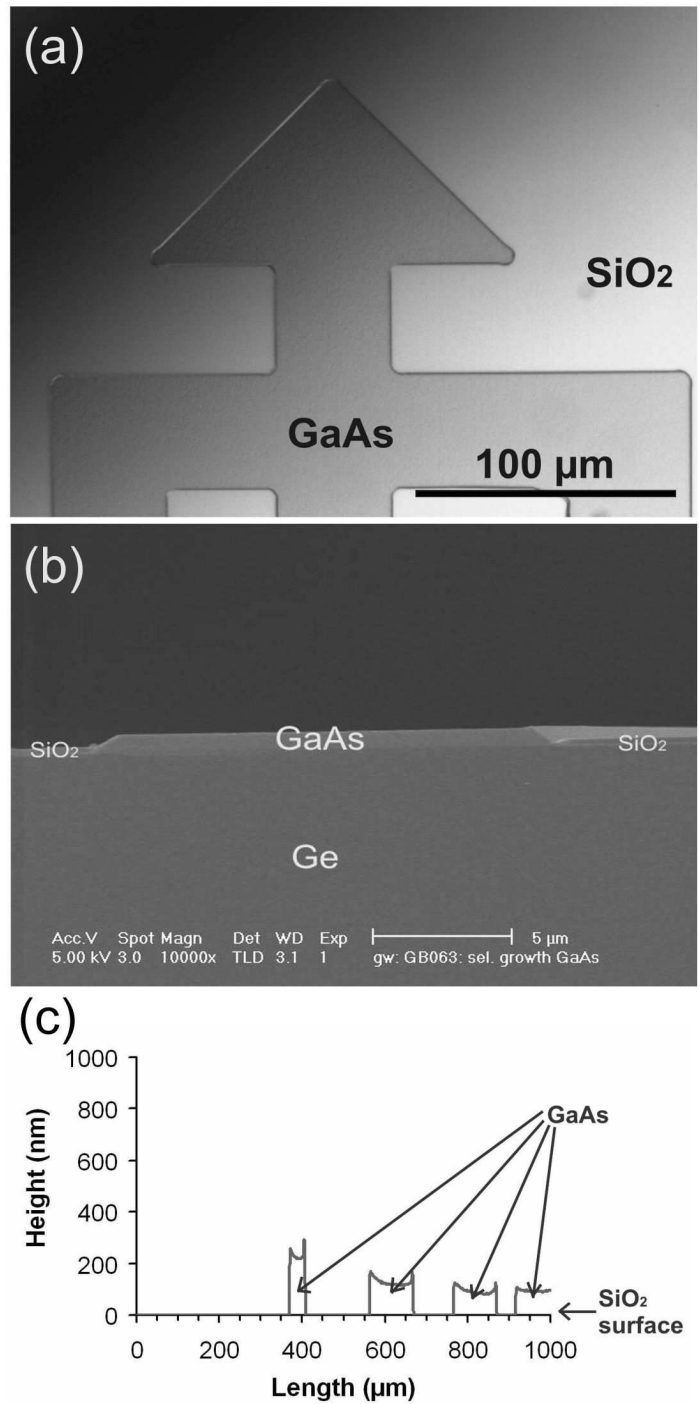

**Figure 5:** Nomarski microscope image (a), cross section SEM picture (b) and profilometer measurement (c) of a selectively grown GaAs layer on Ge, grown with the optimized growth procedure.

## 6. Material characterization

Having shown the macroscopic characteristics of the selectively grown layers, we now want to focus on the microscopic properties of the selectively grown material with x-ray diffraction (XRD) and photoluminescence (PL) spectroscopy. In detail, we are going to compare the selectively grown GaAs to planar GaAs grown on Ge. Figure 6 shows a reciprocal space map of the (115) reflection of a planar GaAs film (a) and a selectively grown GaAs film (b) on Ge. Both GaAs films are fully strained with the in-plane lattice constant of the GaAs being exactly matched to the Ge lattice constant. Note that all selectively grown films show the same XRD spectrum, even the poorest quality film grown entirely at low pressure.

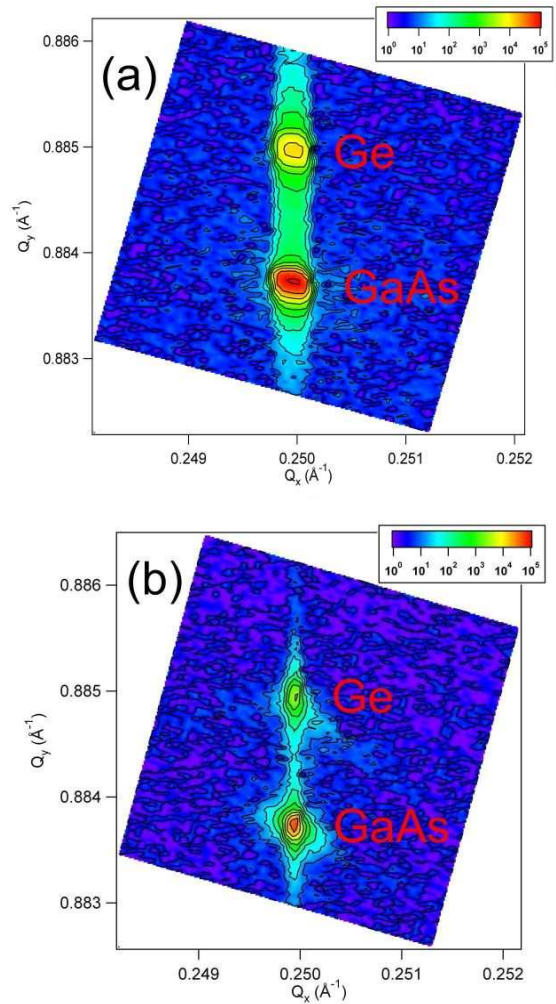

**Figure 6:** Reciprocal space map of the (115) plane x-ray reflection of a planar (a) and a selectively grown (b) GaAs film on Ge.

A much more sensitive technique for GaAs characterization is PL spectroscopy [25]. Figure 7 shows the 77 K PL spectra of the three different selectively grown GaAs films, as well as the PL spectrum of a planar, 1 µm thick GaAs film grown on an unpatterned Ge substrate. The PL spectrum from the high quality planar GaAs film on Ge shows two very narrow peak structures corresponding to the heavy hole and light hole band contributions, which are separated due to small strain in the GaAs layer, arising from the small 0.1% lattice mismatch between Ge and GaAs. The full width at half maximum peak width of the excitonic peaks, which translate the quality of the layer, is equal to 3 meV for the planar GaAs at 77 K, which corresponds to very high quality GaAs films. For more details about the PL structures of thin GaAs on Ge see ref. [25]. Both the selectively grown films with the standard and with the optimized growth procedure show slightly broader PL features. The heavy hole and light hole contribution can not be clearly separated, even though the shape of the spectrum suggests the presence of two overlapping peaks. Nevertheless, despite a probable broadening due to strain variations in the selectively grown layers, the 77 K FWHM energy resolution of the PL peak from the selectively grown structures is still of the order of 10 to 15 meV, which still corresponds to good quality GaAs layers. On the other hand, no PL light can be observed from the low pressure grown film, as the boundaries between APDs act as strong non-radiative recombination centers, inhibiting the emission of any PL radiation.

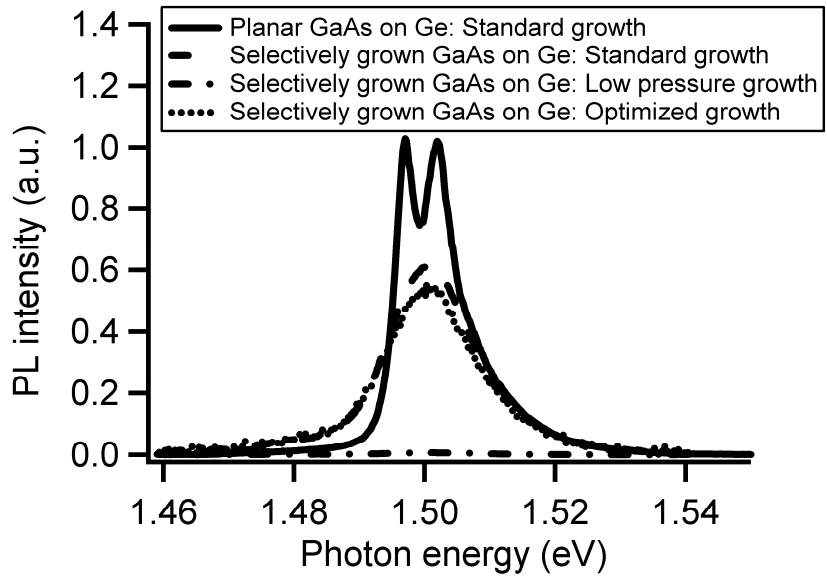

**Figure 7:** 77 K photoluminescence spectra of the selectively grown GaAs films on Ge. The film grown with the standard growth procedure is shown (dashed line) as well as the film grown with the low pressure growth procedure (dashed-dotted line) and with the optimized growth procedure (dotted line). For comparison purposes the spectrum for a planar GaAs film on Ge is shown as well (solid line).

Room temperature PL spectra of the three selectively grown samples are shown in figure 8, along with the room temperature spectrum of a planar GaAs film. Because of reduced lifetime of the charge carriers at room temperature and a consequently reduced diffusion constant, the four structures emit the same spectrum, typical for a lowly doped GaAs layer.

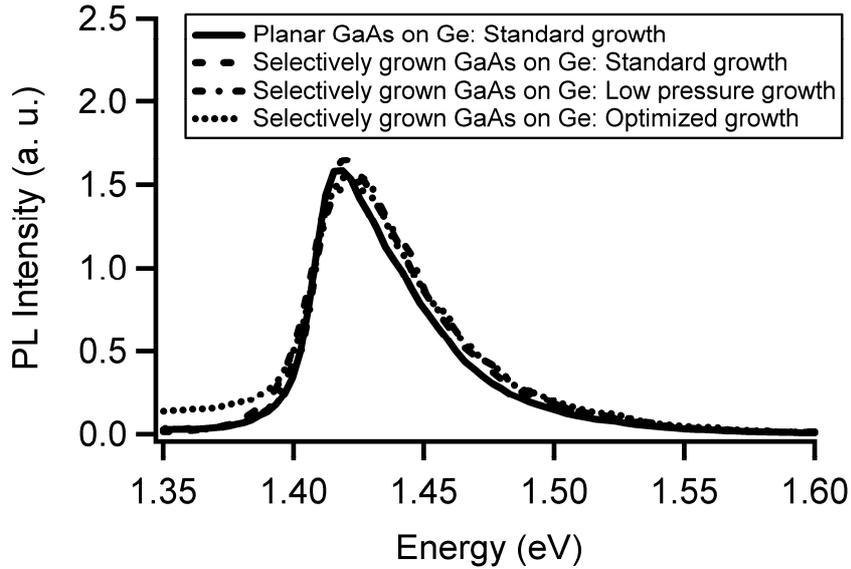

**Figure 8:** Room temperature photoluminescence spectra of the selectively grown GaAs films on Ge. The film grown with the standard growth procedure is shown (dashed line) as well as the film grown with the low pressure growth procedure (dashed-dotted line) and with the optimized growth procedure (dotted line). For comparison purposes the spectrum for a planar GaAs film on Ge is shown as well (solid line).

## 7. Conclusions

We have demonstrated selective growth of thin GaAs films on Ge substrates with a 200 nm thick $SiO_2$ mask layer. An optimized growth procedure was developed, which reduces the thickness of the high pressure nucleation layer to about 13 nm, enabling a good compromise between the inhibition of anti phase domain growth and the reduction of loading effects. Residual loading effects are still present for growth with the optimized growth procedure. 77 K photoluminescence spectroscopy characterization of the selectively grown films shows full width at half maximum peak widths of 10-15 meV, demonstrating the good quality of the selectively grown GaAs on Ge.